\documentclass [a4paper, 11pt] {article}
\usepackage{amscd}
\usepackage{amstex}
\usepackage{epsfig}

\title{How fast do structures emerge in hypercycle-systems?}

\author{Stephan Altmeyer, Claus Wilke, and Thomas Martinetz\\
Institut f\"ur Neuroinformatik\\
Ruhr-Universit\"at Bochum\\ 
D-44780 Bochum, Germany\\ 
Stephan.Altmeyer@@neuroinformatik.ruhr-uni-bochum.de}

\begin{document}
\maketitle

\begin{abstract}
A general framework for the simulation of reaction-diffusion
systems with 
probabilistic cellular automata is presented. The basic reaction probabilities 
of the chemical model translate directly into the transition rules of the 
automaton, thus allowing a clear comparison between simulation results and 
analytic calculations. This framework is then applied to simulations
of hypercycle-systems in up to three dimensions.\\
Furthermore, a new measurement quantity is introduced and applied to the
hypercycle-systems in two and three dimensions. It can be shown that
this quantity can be
interpreted as a measure for the macroscopic order of the hypercycle systems.  
\end{abstract}

%body of the text
\section{The automaton}
In the past a lot of work was done in the field of algorithmic
chemistry \cite{Bagley92,Bagley292} and chemistry on cellular automata
(CA) \cite{Gerhardt89,Boerlijst91}. Our goal is to give a complete
framework which begins with the determination of the transition rules
from basic reaction probabilities and ends with the comparison between the
emergence of spatially ordered structures and a macroscopic measurement
quantity in hypercycle-systems.\\ 
In this paper we can only give a brief introduction of our
probabilistic CA with asynchronous
updating. Our aim was to establish a one-to-one
correspondence between real chemical reactions (defined by their
reaction probabilities) and the simulations on a CA (defined by the
transition rules).
Therefore, we explain the calculation of these transition rules for a simple
particle production process in detail and then present the transition rules for
more complex reaction types without derivation.  
\subsection{Particle production processes}
In our system, particles can be produced through spontaneous or induced
reactions. The result of both reactions is the change of an empty cell
to an occupied one. Since we evaluate production processes for empty
cells, we will usually have the situation that several different
reactions are possible, resulting in a different final state for the
previously empty cell. If we think, e.g.\, of spontaneous
self-replication, all the nearest neighbors of the empty cell are in
principle equally likely to reproduce into this cell. In chemistry such
symmetry breaking phenomena are induced by statistical fluctuations,
e.g.\ temperature gradients.\\
Let $p_i,\; i \in \cal{A}$, be the probability for a single species $i$
to replicate into the empty cell by a special reaction, e.g.\
self-replication. Here $\cal{A}$ denotes an index set containing all indices
which are necessary to enumerate all species in the given context and
$N_i$ will denote the number of molecules of species $i$.\\
In the case there are several nearest neighbor cells
occupied, we first have to calculate the
total probability $p^{\rm tot}$ that at least one arbitrary
nearest-neighbor-molecule will occupy the empty cell. We get
\begin{equation}
p^{\rm tot}=1-\prod_{i \in \cal{A}}(1-p_i)^{N_i} \label{eq.total}.
\end{equation}
From $p^{\rm tot}$ we can calculate the effective probabilities for
the nearest neighbor molecules to replicate in the given arrangement
of species. We will denote these probabilities with $p_i^{\rm
  eff}$. In this calculation we have to make sure that the ratio
$p_i/p_j$ is equal to $p_i^{\rm eff}/p_j^{\rm eff}$,
$\forall i,j \in \cal{A}$.  Since the sum over all $p_i$ is not
necessarily 1, we introduce a normalization factor $\sum_{i
  \in \cal{A}}N_ip_i$.\footnote{The normalization factor is introduced
  to ensure that the sum over all $p_i^{\rm eff}$ and $1-p^{\rm tot}$
  is 1.} 
We get
\begin{equation}
p_i^{\rm eff}=p_i \frac{p^{\rm tot}}{\sum_{i \in \cal{A}}N_ip_i}.
\end{equation}
Note that $p^{\rm tot}$ is the sum of $N_ip_i^{\rm eff}$
over all $i$.
\subsubsection{Type I reactions}
Reactions of type I,
\[M_i@>{p_{i,m,n}}>>M_m + M_n,\]
represent the group of spontaneous reactions with
two reaction products. After some work and with the calculations done
above we get:
\begin{equation}
p_{i,m,n}^{\rm eff}=p_{i,m,n} \frac{1-\prod_{i \in
    \cal{A}}(1-p_{i,m,n})^{N_i}}{\sum_{i \in \cal{A}}N_i p_{i,m,n}}.
\end{equation}
The  corresponding transition rule then reads:\\
{\em A cell with state zero will change to state $n$ and a
  nearest-neighbor cell with state $i$ will change to $m$ with the
  probability $p_{i,m,n}^{\rm eff}$.}
\subsubsection{Type II reactions}
Reactions of type II,
\[M_i + M_j@>{p_{i,j,m,n,o}}>>M_m + M_n + M_o,\]
describe a group of two-body reactions with three
reaction products. Obviously, such reaction processes are more difficult to
handle since we have to implement the determination of the correct
reaction partner. The related quantities are primed.\\
After some calculations we get:
 \begin{equation}
p_{i,j,m,n,o}^{\rm eff}=p'_{i,j,m,n,o} \frac{1-\prod_{i \in
    \cal{A}}(1-p'_{i,j,m,n,o})^{N'_i}}{\sum_{i \in
    \cal{A}}N_i p'_{i,j,m,n,o}}. 
\end{equation}
with
\begin{equation}
p'_{i,j,m,n,o}=p(N' \neq 0)\frac{N'_j}{N'}p_{i,j,m,n,o}.
\end{equation}
$p(N' \neq 0)$ denotes the probability that there is at least one
molecule to react with. For the derivation of $p'_{i,j,m,n,o}$ we have
implicitly used the idea of an attractive force
between the molecules which means that the reactive molecule therefore "jumps"
to a reaction partner. On the other hand, if we do not assume such an
attractive force, which means a molecule can also "jump" into a hole, we
have to substitute $p'_{i,j,m,n,o}$ with $N'_j/N'_{\rm max}$.\\
Then we can write down the transition rule:\\
{\em The probability that a cell in state zero, with a nearest-neighbor
  cell in state $i$ and with a "reaction" cell in state $j$ will change
  to $o$, while the nearest-neighbor cell changes to $m$ and the
  "reaction" cell changes to $n$, is $p_{i,j,m,n,o}^{\rm eff}$.}
\subsubsection{Particle decay}
The decay process, 
\[M_i@>{p^{\rm dec}_i}>>0,\]
can change an occupied cell into an empty one. The transition rule
description is rather easy:\\
{\em The probability that a cell in state i will change to a cell in
  state zero is $p_i^{\rm dec}$.}
\subsection{Diffusion}
The diffusion is an additional process. Since we use an
asynchronous automaton, an often used diffusion algorithm for CA
\cite{Toffoli87} is not necessary. Instead, we use a more natural
algorithm that can be divided into three subprocesses:
\begin{enumerate}
\item An arbitrary cell of the CA is chosen.
\item One of it's nearest neighbor cells is determined at random.
\item The states of the two cells are interchanged.
\end{enumerate}    
\subsection{Automaton update}
We use asynchronous updating, since this seems to be the more natural
way \cite{Harvey97}. The basic processes are then executed in the
following way:\\
Choose an arbitrary cell at random. If the cell is occupied we use the
decay-transition rule. Otherwise the transition rule for
reactions of type I or of type II or a combination of both are used.\\
The reaction processes and the diffusion process are then combined to the
full simulation in the following way. Let $k$ be the total number of
cells in a CA and $d$ the
ratio of the number of basic processes and the number of diffusion
steps. Then a  CA-time-step $t$ consists of $k$ basic processes and $dk$
diffusion steps. It is important that the basic processes and the
diffusion steps are mixed up very well.\\
Note that $t$ is independent of the system's size, since we scale the
number internal processes with it.
\section{Simulation}
We have simulated systems with (symmetrical) hypercycles
\cite{Eigen79,Hofbauer88}, which means we set
\begin{equation}
p_{i,j,k}=
\begin{cases}
  p^{\rm self} & \text{if $i=j=k$},\\
  0 & \text{else}
\end{cases} 
\end{equation}
and\\
\begin{equation}
p_{i,j,m,n,o}=
\begin{cases}
  p^{\rm cat} & \text{if $i=m=o, j=n$ and $i=j+1$ (cyclic)},\\
  0 & \text{else}
\end{cases}
\end{equation}
\subsection{Stable patterns}
\begin{figure}
\centerline{
        \epsfxsize=0.45\columnwidth{\epsfbox{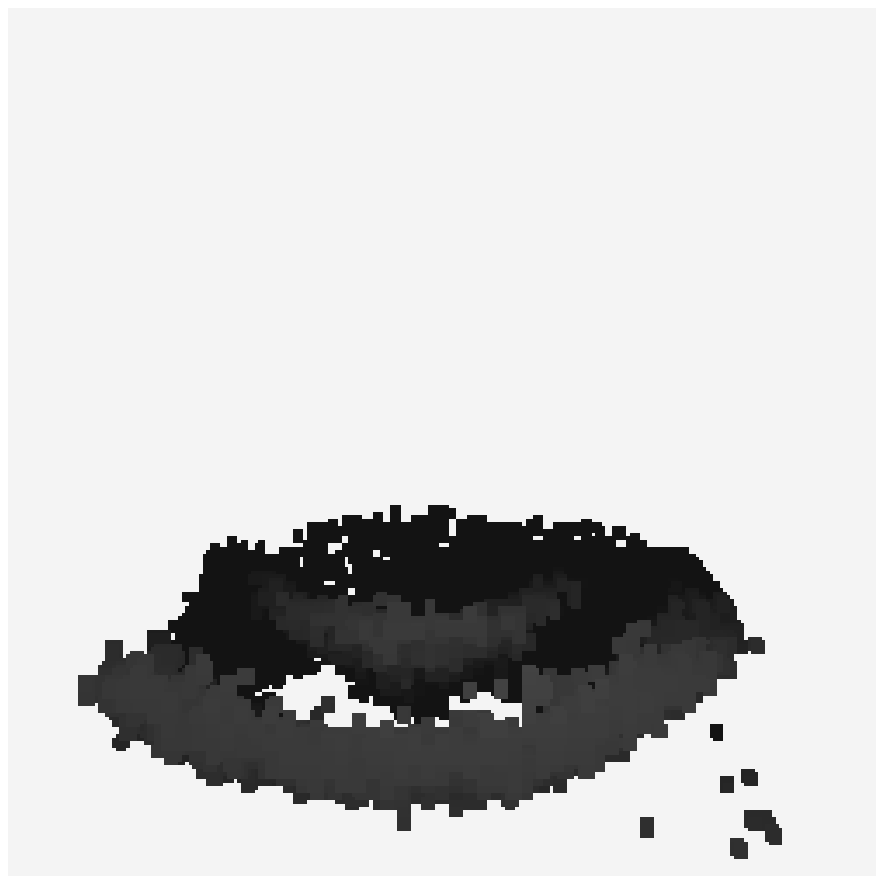}}\hspace{0.05\columnwidth}{\epsfbox{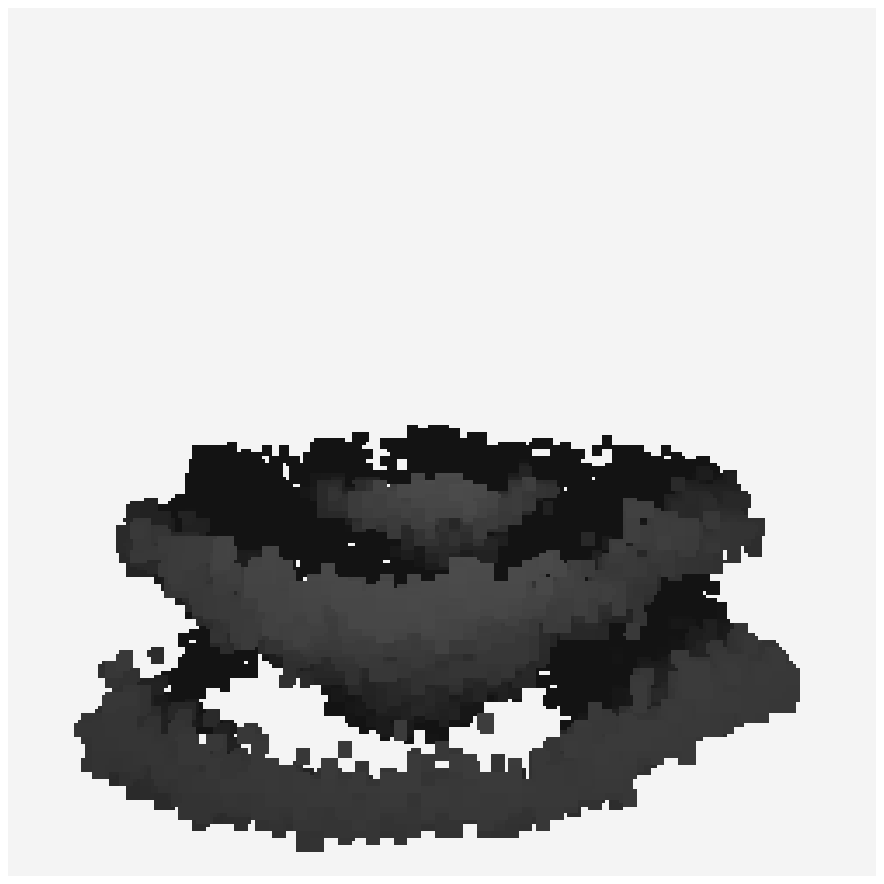}}}
\caption{\label{fig:3d1}The first two pictures show the system's behavior shortly
        after the "injection" of the 2d spiral. We can see the
        emergence of a rotating screw. The left picture was taken
        at $t=150$ and the right one at $t=250$.} 
\end{figure}

\begin{figure}
\centerline{
        \epsfxsize=0.45\columnwidth{\epsfbox{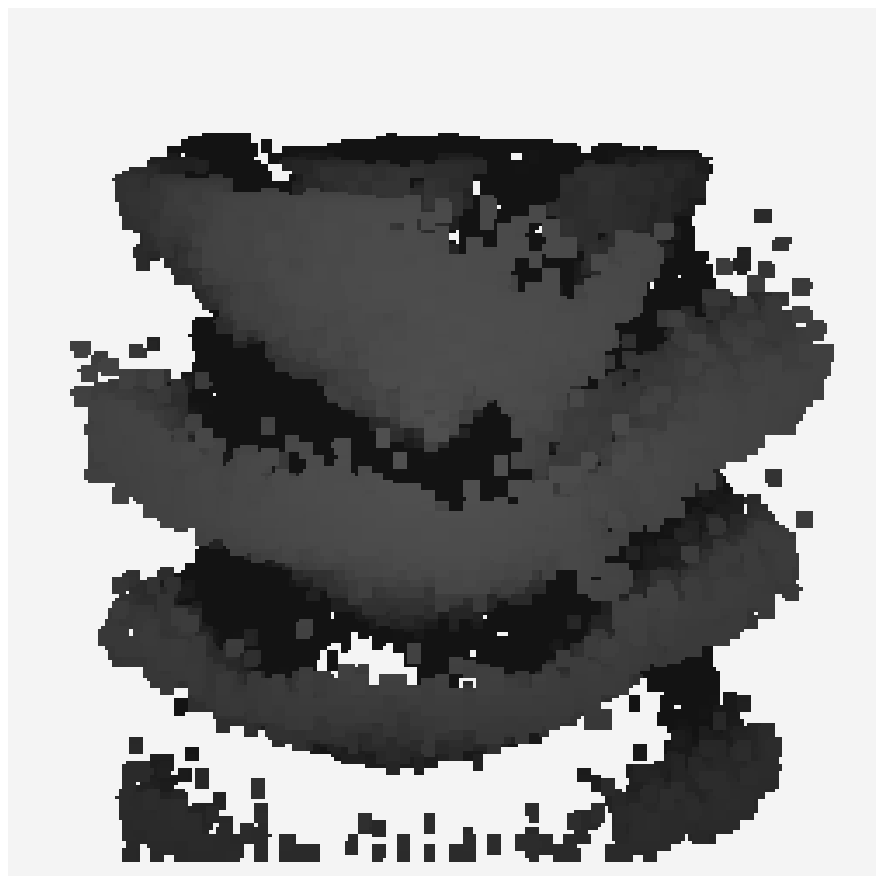}}\hspace{0.05\columnwidth}{\epsfbox{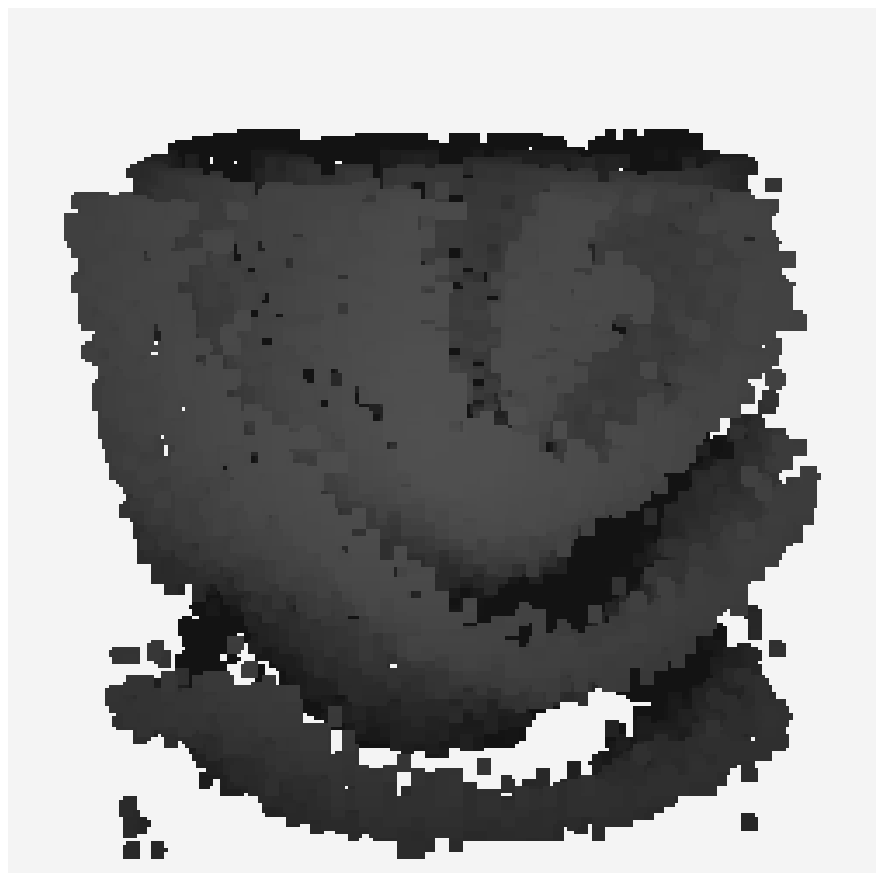}}}
\caption{\label{fig:3d2}If the space is nearly completely occupied the molecules
        cannot replicate freely and have to compete for empty
        space. Therefore the structure changes from a screw to a
        cone-like pattern.}
\end{figure}

In this part of the paper we will show the well-known
spiral waves \cite{Boerlijst91} in two dimensions and the emergence of
a stable scroll wave structures \cite{Tyson88} in three
dimensions. In the latter case we have used a two-dimensional spiral
as initial condition. We will see the system going through
different unstable patterns. For reasons
of clear visualization, in
the pictures we show only one of the six species present in the simulation.\\
The three-dimensional analog of a spiral wave is a scroll wave \cite{Tyson88}.
As we can see in Fig.~\ref{fig:2d2}, even in the two dimensional
case, structures can be very complex. Since there are more degrees
of freedom in three spatial dimensions, the structures arising from
non-ordered initial conditions are much more complex. Therefore, we use
a 2d spiral as a symmetry-breaking initial condition for our
simulations to simplify the emergent structures.\\
We can see the spiral's growth
in the z-direction (see Fig.~\ref{fig:3d1},~\ref{fig:3d2}). As a matter of fact, the reproduction activity of a
species becomes high if it is adjacent to it's "supporter".   
Therefore, in a layer above, molecules are first created over the small
region where their own and their "supporter" are present. Since the
spirals rotate in the "supporter" direction, we can see the
phenomenon that the new spirals, growing on top of the old ones, are
phase shifted in the rotation direction. This continues until the
space in z-direction
is occupied. The resulting structure looks like a rotating
screw. When all the available space is occupied, the molecules cannot
replicate 
freely any more. Since space is rare than, the molecules have to compete for empty
space to replicate. For this reason, the molecules begin to
arrange themselves in a
different way so that the "free" screw begins to change to a
cone structure. This cone structure is stable so that a real
scroll wave cannot be seen.\\ 
The whole re-arrangement process takes much more time than the
occupation at the beginning. The pictures show the emergence of the cone
structure. At the top of the cone the emergence of other
small cones can be seen so that the whole structure becomes more complex.\\

\subsection{A microscopic measurement quantity}
Now we will introduce a measurement quantity with which we can measure
the arrangement of the molecules in our automaton. The definition of
this order-quantity, called $\alpha_k$, is rather easy:\\
For every molecule, we take the nearest neighbor molecules. If an adjacent
molecule is a "supporter", the variable $S$ is incremented by
one. Otherwise the variable $N$ is incremented by one. If a cell is
not occupied, nothing is done. With these microscopic values we can
define the macroscopic quantity $\alpha_k$:
\begin{equation}
\alpha_k = \frac{kS}{N+kS},
\end{equation}
where $k \geq 0$ is a tuning constant. The larger $k$, the more
sensitive is $\alpha_k$ on $S$.\\
Since both $N,S \geq 0$, the quantity $\alpha_k$ ranges between 0 and 1. We will
later see that $\alpha_k$ is even constant for large $t$. (Obviously,
if $\alpha_k$ is constant, $\alpha_{k'}$, with another $k' \neq k$,
is also constant.)\\ 
Moreover, since we do not take into account empty cells, $\alpha_k$ is
independent
of the system's concentration and therefore independent of $p^{\rm
  dec}$.\\
For our simulations we have used $k=10$ and applied $\alpha_{10}$ to
both the two and three dimensional case.\\
Note that there are a lot of possible microscopic/macroscopic
quantities which all can describe the system from different points of view.
\subsection{Two dimensions}
For the two dimensional case we have used two different initial
conditions. In both cases the parameter set is $p^{\rm dec}=0.1$,
$p^{\rm self}=0.05$, $p^{\rm cat}=0.9$ and $d=1$.\\ 
\begin{figure}
\centerline{
        \epsfxsize=0.3\columnwidth{\epsfbox{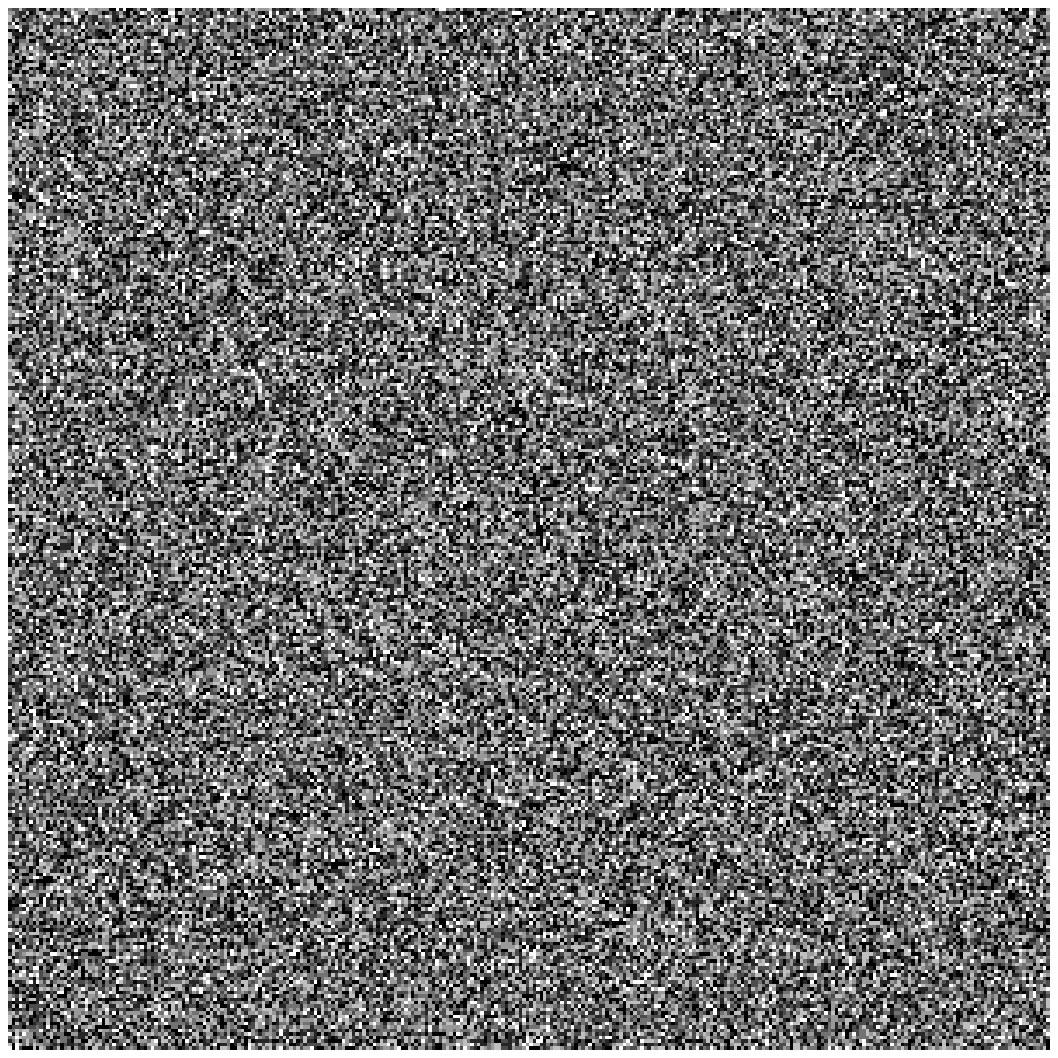}}\hspace{0.05\columnwidth}{\epsfbox{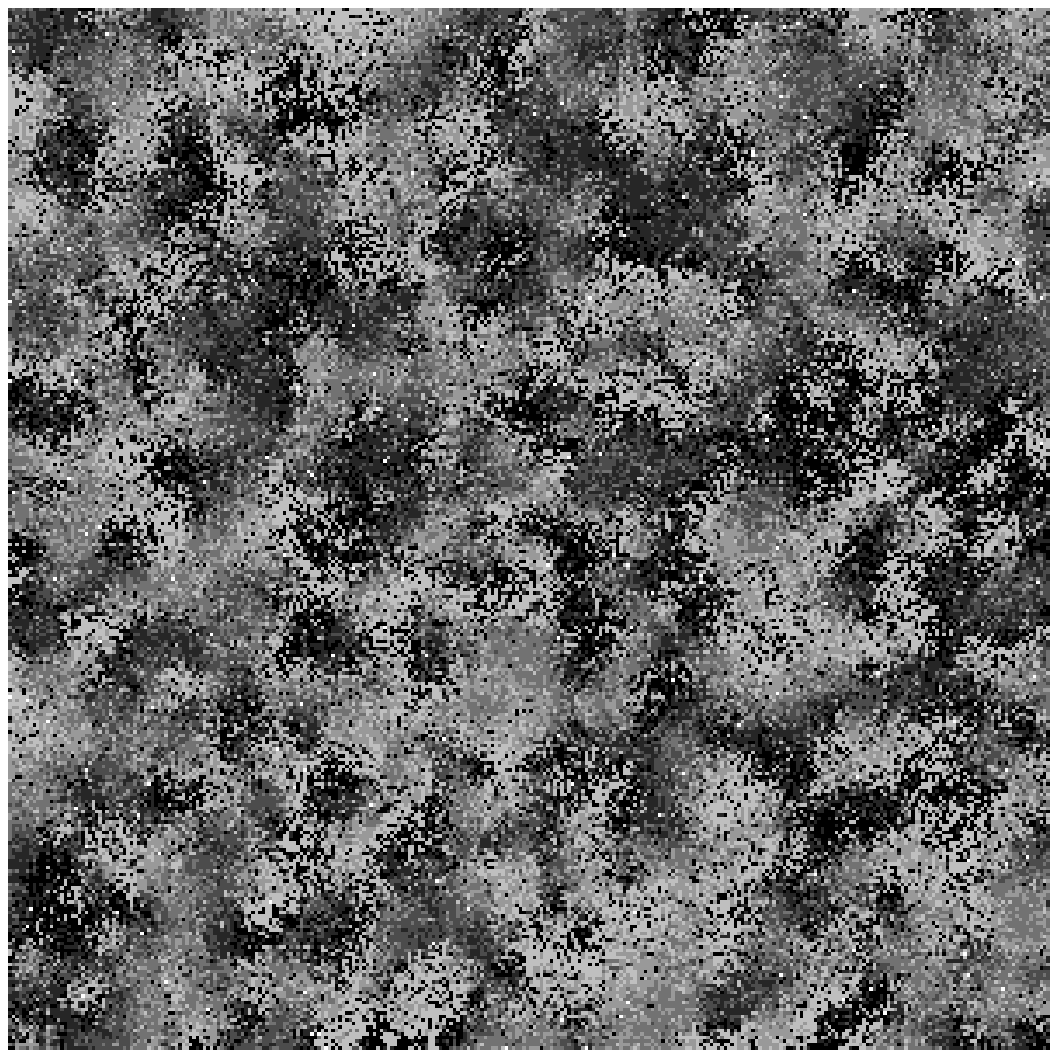}}\hspace{0.05\columnwidth}{\epsfbox{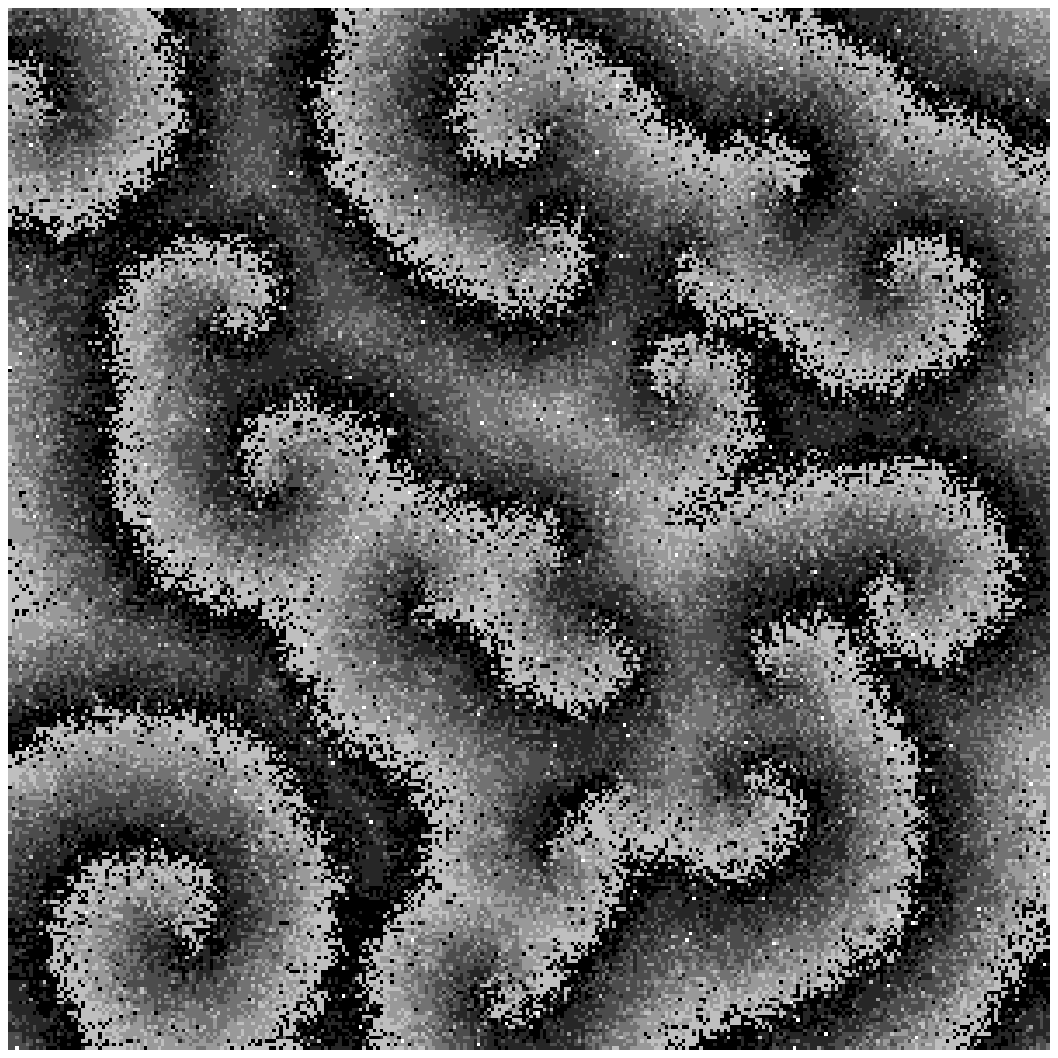}}}
\caption{\label{fig:2d2}The three pictures show the time dependent
        behavior of the 
        system. The times when the pictures have been taken are (from
        left to right): $t=0$, $t=120$ and $t=9000$. See also
        Fig.~\ref{fig:2d1} and text for further information.}
\end{figure}
\begin{figure}
\centerline{
        \epsfxsize=0.8\columnwidth{\epsfbox{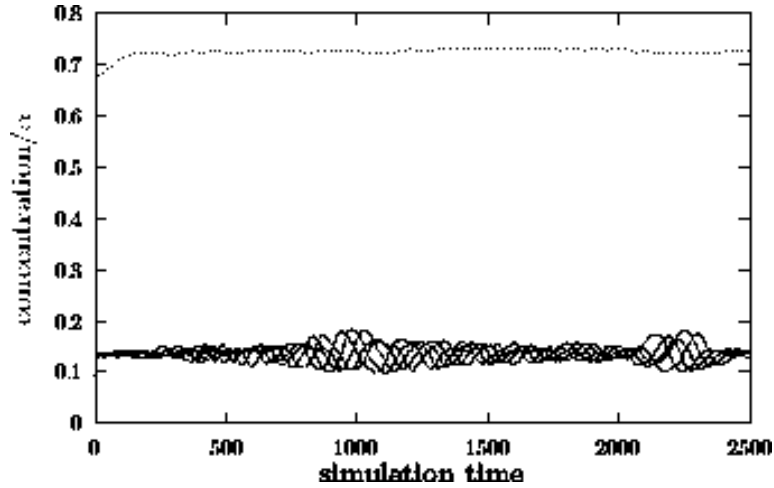}}}
\caption{\label{fig:2d1}The picture shows the concentrations of the
        six hypercycle
        species (solid lines) and the value of $\alpha_{10}$ versus
        time for a system with random initial condition.}
\end{figure}

First, the hypercycle-system was initialized at
random, see Fig.~\ref{fig:2d2}. We have seen $\alpha_{10}$
converging rapidly to a constant value at the very beginning. After
$t=120$, however, $\alpha_{10}$ changes only in very small amounts. In
comparison, the global structure of the system also changes very slowly,
see Fig.~\ref{fig:2d1}. It seems that the system's rough structure
is determined at the very beginning and will then only be
refined, later on. This is very interesting, since there are a
lot of possible stable patterns. However, the decision is made very
early.\\
For the asymptotic value of $\alpha_{10}$ we have found $\alpha_{10}
\approx 0.72$, independent of the number of species in the hypercycle
and of the system's size. Furthermore, $\alpha_{k}$ depends only weakly
on $d$ for $d < 1$.
\begin{figure}
\centerline{
        \epsfxsize=0.3\columnwidth{\epsfbox{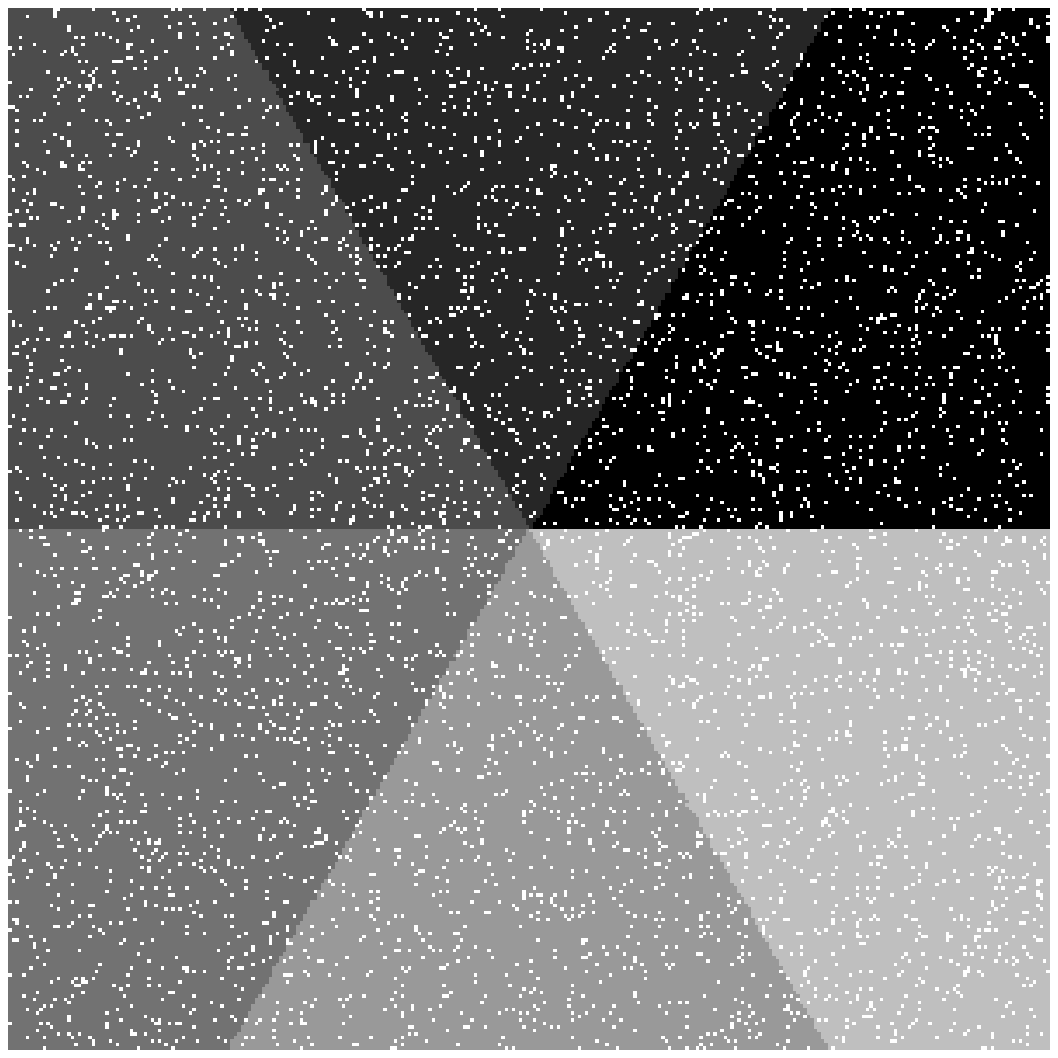}}\hspace{0.05\columnwidth}{\epsfbox{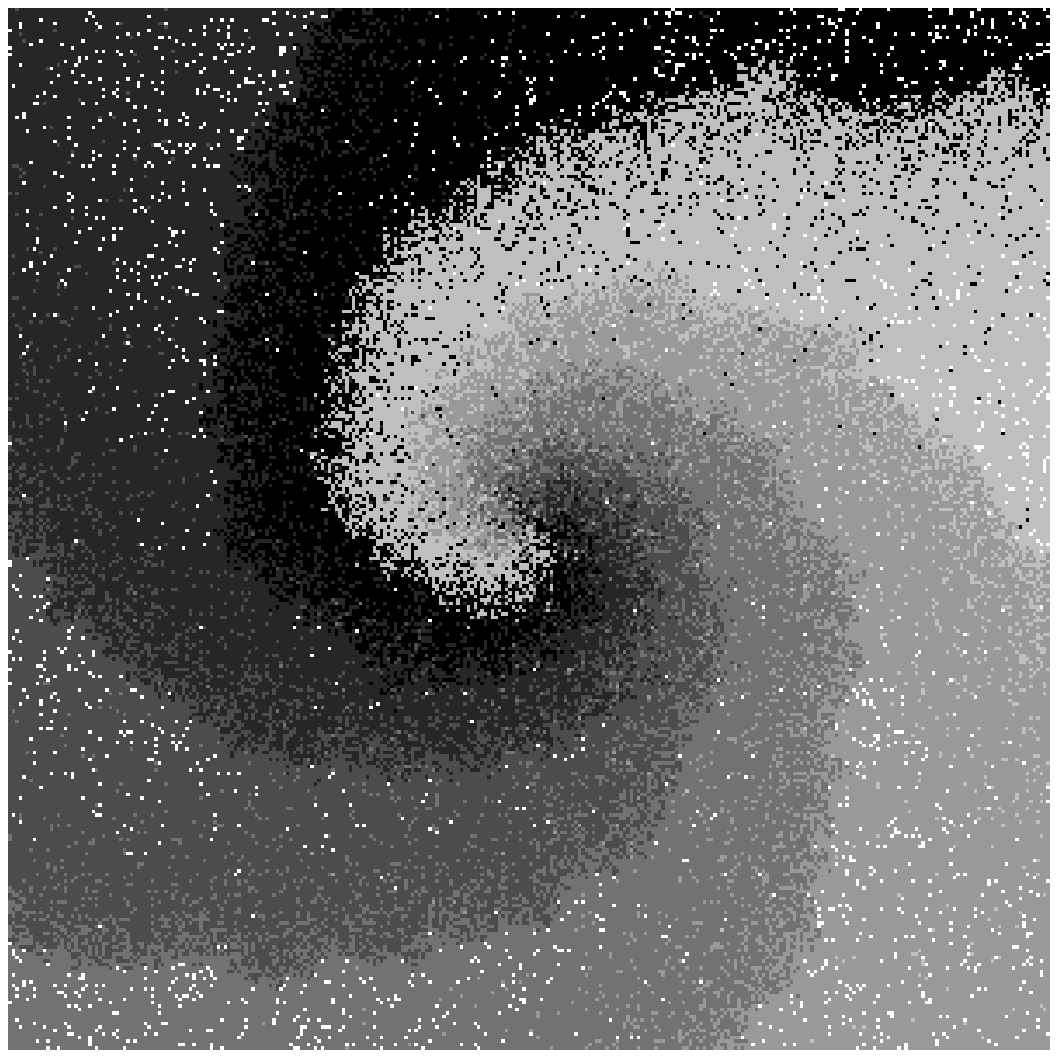}}\hspace{0.05\columnwidth}{\epsfbox{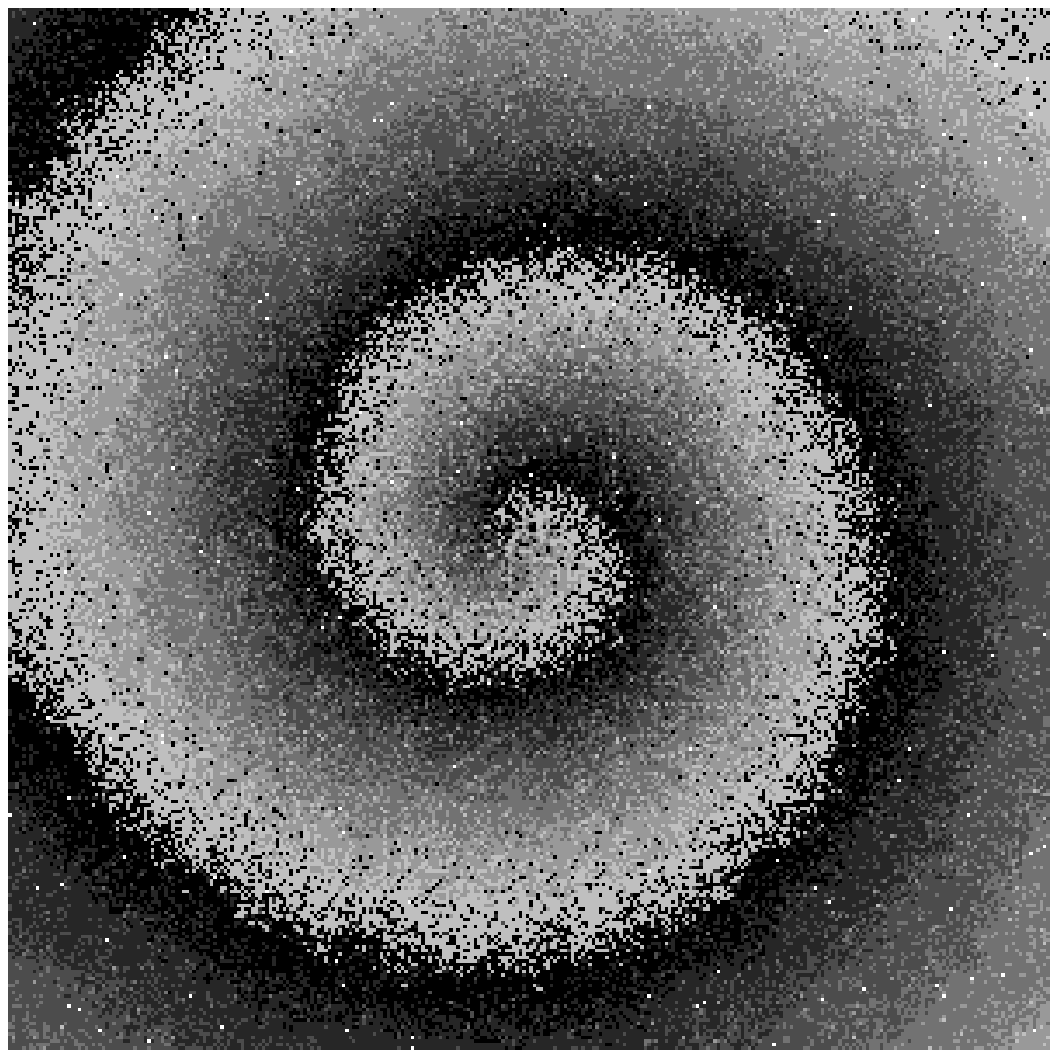}}}
\caption{\label{fig:2d4}The three pictures show the time dependent behavior of the
        system. The times when the pictures have been taken are (from
        left to right): $t=0$, $t=140$ and $t=550$. See also
        Fig.~\ref{fig:2d3} and text for further information.}
\end{figure}
\begin{figure}
\centerline{
        \epsfxsize=0.8\columnwidth{\epsfbox{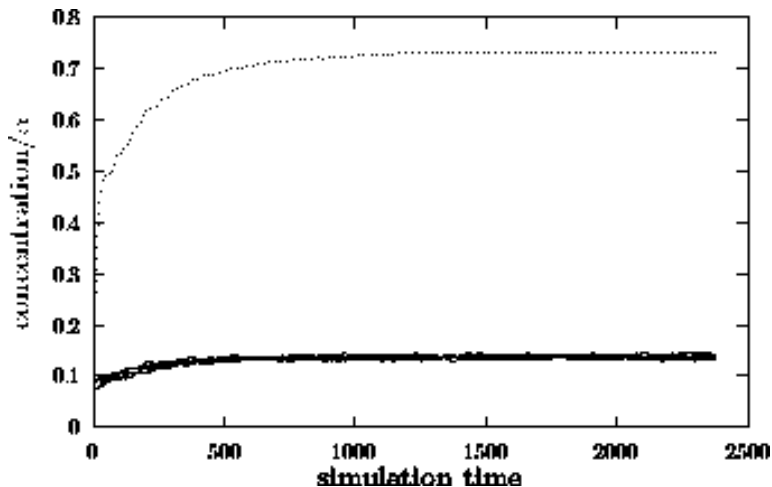}}}
\caption{\label{fig:2d3}The picture shows the concentrations of the six hypercycle
        species (solid lines) and the value of $\alpha_{10}$ versus
        time for a system with ordered triangles as initial condition.}
\end{figure}
The next step was to examine a system where only one stable
final pattern exits. Therefore, we have used a special initial
condition which is displayed in Fig.~\ref{fig:2d4}. The final structure will be
of course a single rotating spiral.\\
What we have seen was $\alpha_{10}$ changing comparably to the
transformation of the triangular structure
to the spiral structure. Since the major part of this transformation
is done at the beginning, $\alpha$ also increases fast during the first
simulation steps and then increases slower and slower. The system
behaves similar to the case with random initialization, although the
convergence of $\alpha_{k}$ to it's asymptotic value is significantly slower. 
But what is more surprising, the asymptotic value of $\alpha_{10}$ is
again the same, $\alpha_{10} \approx 0.72$.\\
We will now go a step beyond and apply
$\alpha_{10}$ on our three dimensional simulator. 
\subsection{Three dimensions}
\begin{figure}
\centerline{
        \epsfxsize=0.8\columnwidth{\epsfbox{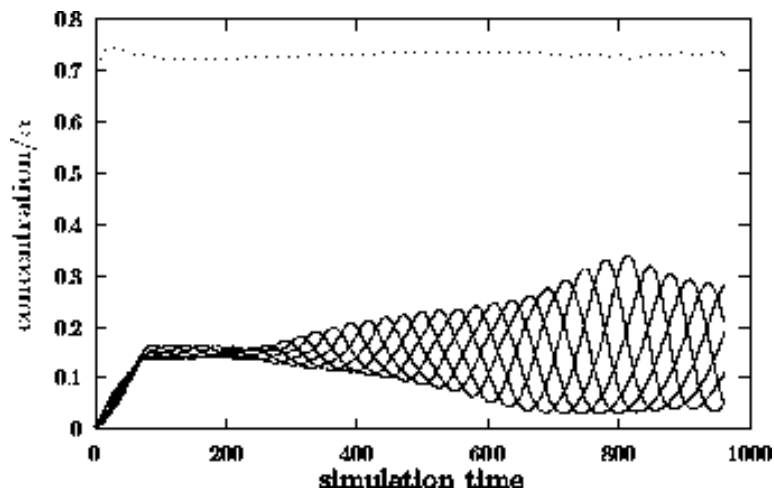}}}
\caption{\label{fig:3dgraph}The picture shows the concentrations of
        the six hypercycle 
        species (solid lines) and the value of $\alpha_{10}$ versus
        time for a system in three spatial dimensions. Note that a single two
        dimensional spiral was used as initial condition.}
\end{figure}
As we have seen above (Fig.~\ref{fig:3d1},~\ref{fig:3d2})
in three dimensions there are several different unstable structures,
beginning with a single two dimensional spiral as initial condition
and ending with a stable cone-structure.\\ 
In Fig.~\ref{fig:3dgraph} we can see $\alpha_{10}$ starting with
$\alpha_{10} \approx 0.72$. But then we see $\alpha_{k}$ increasing in
the time interval when the spiral occupies the empty space by growing
in the z-direction. After that, when the available space is occupied,
new boundaries are present and $\alpha_{10}$ decreases to the former value.\\
This is a bit surprising since the cone structures are quite
different and at the first glance there seems to be no reason why the
macroscopic order should be similar.
\section{Conclusions}
In this paper we have shown pictures of the transformation from two
dimensional patterns
to three dimensional ones and have therefore seen the system
walking through different states of arrangement. But what is more
interesting is the fact that our macroscopic quantity $\alpha_{k}$
(here: $\alpha_{10}$) is quite constant after a short time and has
even the same asymptotic value in different settings.\\
In all our simulations, especially in two dimensions, we have seen
$\alpha_{10}$ converging rapidly to the fixed value, $\alpha_{10}
\approx 0.72$. Thereof we can deduce that such catalytic systems have
the urge to determine their final structures at the very
beginning. Later on structures are only refined.\\ 
Although the hypercycles have been studied very well in the past years
some new features can still be found. We think that the definition of
microscopic quantities and their interpretation in a macroscopic sense
is one of the right ways to learn more about catalytic systems which
can be even more complex than such "simple" symmetric hypercycles.
\section{Outlook}
It seems to be interesting for future work to study whether stability
against parasites is possible in three dimensions, similar to the
two-dimensional case \cite{Boerlijst91,Cronhjort96}. Furthermore, the
behavior of $\alpha_{k}$ in such parasite-systems might be interesting.

\end{document}